\documentclass[sigconf,screen]{acmart}

\AtBeginDocument{%
 }

\usepackage{graphicx}
\usepackage[utf8]{inputenc}
\usepackage{tabularx}
\usepackage{comment}
\usepackage{booktabs}
\usepackage{soul}
\usepackage{xcolor}
\usepackage{xspace}
\usepackage{adjustbox}
\usepackage{float}
\usepackage{textcomp}
\usepackage{cleveref}
\usepackage{listings}
\usepackage[noend]{algpseudocode}
\usepackage{balance}
\usepackage{multirow}
\usepackage{colortbl}
\usepackage{threeparttable}
\usepackage{multicol}
\usepackage{tcolorbox}
\usepackage{soul}
\usepackage{longtable}
\usepackage{array}
\usepackage{pdflscape} 
\usepackage{fancyhdr} 
\usepackage{rotating}
\usepackage{url} 
\usepackage{multirow}
\usepackage{pifont}
\usepackage{todonotes}

\definecolor{jbblue}{HTML}{0573EB}
\definecolor{jbgreen}{HTML}{1D8574}
\definecolor{jborange}{HTML}{C15703}
\definecolor{jbpurple}{HTML}{6b57ff}
\definecolor{jbpink}{HTML}{E90067}
\definecolor{lightgray}{gray}{0.9} 
\definecolor{codegreen}{rgb}{0,0.6,0}
\definecolor{codegray}{rgb}{0.5,0.5,0.5}
\definecolor{codepurple}{rgb}{0.58,0,0.82}
\definecolor{backcolour}{rgb}{0.95,0.95,0.92}
\definecolor{APA_stats}{RGB}{100, 100, 120}
\definecolor{main}{HTML}{5989cf}  
\definecolor{sub}{HTML}{cde4ff}   
\colorlet{shadecolor}{sub}
\newcommand{\blue}[1]{{\sethlcolor{jbblue!50}\hl{#1}}}
\newcommand{\orange}[1]{{\sethlcolor{jborange!50!}\hl{#1}}}
\newcommand{\pink}[1]{{\sethlcolor{jbpink!50!}\hl{#1}}}
\newcommand{\purple}[1]{{\sethlcolor{jbpurple!50!}\hl{#1}}}
\definecolor{mygray}{rgb}{0.8,0.8,0.8}
\usepackage{realboxes}
\DeclareDocumentCommand{\clist}{v}{%
  \Colorbox{mygray}{\csname lstinline\endcsname!#1!}%
}

\newcommand{\eg}{\emph{e.g.,}\xspace}
\newcommand{\nSurvey}{56\xspace}
\newcommand{\nDesign}{7\xspace}
\newcommand{\nConfigs}{33\xspace}

\usepackage[english]{babel}
\usepackage[autostyle, english = american]{csquotes}
\MakeOuterQuote{"}

\usepackage{tikz}
\newcommand{\circled}[1]{\tikz[baseline=(char.base)]{\node[shape=circle,draw,inner sep=1pt] (char) {#1};}}

\newcommand{\company}{JetBrains\xspace}

\copyrightyear{2026}
\acmYear{2026}
\setcopyright{cc}
\setcctype{by}
\acmConference[FSE Companion '26]{34th ACM Joint European Software Engineering Conference and Symposium on the Foundations of Software Engineering}{July 05--09, 2026}{Montreal, QC, Canada}
\acmBooktitle{34th ACM Joint European Software Engineering Conference and Symposium on the Foundations of Software Engineering (FSE Companion '26), July 05--09, 2026, Montreal, QC, Canada}
\acmDOI{10.1145/3803437.3806709}
\acmISBN{979-8-4007-2636-1/2026/07}

\begin{document}

\title{Configurable AI Coding Assistants:\\Designing For Developers Who Like to Be in Control}
\renewcommand{\shorttitle}{Configurable AI Coding Assistants}
\author{Ekaterina Koshchenko}
\authornote{Both authors contributed equally to this research.}
\affiliation{%
 \institution{JetBrains Research}
 \city{Amsterdam}
 \country{Netherlands}
}
\email{ekaterina.koshchenko@jetbrains.com}

\author{Jovana Stankovic}
\authornotemark[1]
\affiliation{%
 \institution{JetBrains Research}
 \city{Belgrade}
 \country{Serbia}
}
\email{jovanas769@gmail.com}

\author{Ilya Zakharov}

\affiliation{%
 \institution{JetBrains Research}
 \city{Belgrade}
 \country{Serbia}
}
\email{ilia.zaharov@jetbrains.com}

\author{Agnia Sergeyuk}
\affiliation{%
 \institution{JetBrains Research}
 \city{Belgrade}
 \country{Serbia}
}
\affiliation{%
 \institution{Delft University of Technology}
 \city{Delft}
 \country{Netherlands}
}
 \email{agnia.sergeyuk@jetbrains.com}

\date{January 2026}

\begin{abstract}
AI coding assistants are now widely used in professional development, yet they offer only limited ways for developers to control how they behave.
In this paper, we investigate what kinds of configurations experienced developers want in coding assistants, how they prioritize different types of configuration needs, and which interface mechanisms they prefer.
We first synthesize product documentation and prior research on trust and personalization to compile a list of \nConfigs configuration options, grouped into four categories: Code suggestions, System \& policies, Human–assistant interaction, and Users \& their personal context. 
We then conduct a survey with \nSurvey professional developers and \nDesign design sessions in which participants arrange configurations into their perfect control board and talk about their needs and experiences in more depth.

Developers report strong interest in configurability: $72.6\%$ of usefulness ratings are positive, while only around a third indicate that the corresponding configuration is known to participants in their tools.
Demand is particularly high for task-related controls such as minimum confidence thresholds, visibility of suggestion quality, and response length, whereas many persona-related configurations are seen as unnecessary.
In this paper, we discuss the implications for designing more unified and discoverable configuration surfaces for future coding assistants.
\end{abstract}



\maketitle

\section{Introduction}

Artificial Intelligence (AI) -powered tools, and AI coding assistants specifically, are rapidly becoming part of everyday software development.
From smart autocompletion and code generation to conversational tools that help with debugging, refactoring, or documentation, developers now use these systems at nearly every stage of the workflow: from early ideation to testing and maintenance.
Surveys show that most developers have already integrated AI tools into their daily work~\cite{DevEco2025, StackoverflowDevSurv2025}.
As these tools become more embedded in development practice, expectations are changing.
Rather than serving as one-size-fits-all helpers, 
there is a growing expectation for personalization and configurability~\cite{HAXdesignspace}.
Developers now want tools that reflect their own working styles, preferred technologies, team practices, and even communication tone.
Some care about how autonomous or directive the assistant feels, others prefer a particular level of verbosity or humanized behavior.

While interest in configurable coding assistants continues to grow, the actual configuration options available today remain limited.
Mainstream tools like GitHub Copilot~\cite{copilot} and Cursor~\cite{cursor}, for instance, offer impressive code completion capabilities but provide little control over aspects such as interaction style.
More broadly, discussions around configurable AI suggest that major vendors are beginning to treat it as a strategic priority.
Yet when it comes to coding assistants in particular, we still know relatively little about what kinds of configurations developers actually want and how they prefer to control them.

Our paper seeks to address the current gap through structured user research with experienced software developers.
We compile a list of \nConfigs configurations (aka "controls") that currently exist in the most popular coding AI assistants, or were suggested in personalization and coding personalization research.
We categorize the identified configurations into broader themes based on what they characterize: \textbf{\textit{\blue{Code suggestions}, \orange{System \& policies}, \pink{Human-Assistant interaction}}}, and \textbf{\textit{\purple{Users \& their personal context}}}.
This categorization helps us to 
uncover which aspects of configurability developers value most and what motivates these preferences, whether it is trust calibration, workflow alignment, or organizational policy.
We also explore how developers wish to express control, by examining preferred interface forms such as toggles, checklists, configuration files, or natural language commands.

Specifically, we aim at answering the following \textbf{Research Questions} (RQs):
\begin{itemize}
  \item[\textbf{RQ1}] What configurations do experienced developers \textbf{want} in AI-assisted coding tools?
  \item[\textbf{RQ2}] How do developers \textbf{prioritize} different types of configuration needs?
  \item[\textbf{RQ3}] What types of configuration \textbf{interfaces}
  do developers prefer to control their AI-assisted coding tools?
\end{itemize}

To answer these questions, we conduct a two-stage study.
First, we survey developers to identify and rank the configuration options they care about most.
Then, we run \nDesign co-design sessions and interviews to dig deeper into developers' preferred settings, interface ideas, motivations, and constraints.
Our findings suggest that developers primarily value configurations that shape what the assistant outputs and how it behaves in context, rather than those that set up rich personal profiles.
They also clearly distinguish between stable, project-level defaults and quick, task-specific tweaks, and that distinction influences both how often they adjust a setting and where they expect to find it in the IDE.

\section{Background}\label{back}
\subsection{Industry overview}\label{back:industry}
AI-powered coding assistants are now a routine part of modern software development~\cite{DevEco2025}.
Tools such as GitHub Copilot~\cite{copilot}, OpenAI ChatGPT~\cite{chatgpt}, Google Gemini~\cite{gemini}, and JetBrains AI Assistant~\cite{jbai} are widely used for code completion, documentation, testing, debugging, and other tasks.
Newer platforms, such as Cursor~\cite{cursor}, are also emerging, following the direction of agent-mode workflows and offering more integrated AI support across the software engineering lifecycle.
In this section, we overview the current state of configurability in five of the most popular coding AI tools according to the State of Developer Ecosystem in 2025 Report~\cite{DevEco2025}.

\textbf{\textit{GitHub Copilot.~\cite{copilot}}}
Copilot offers configuration options spread across GitHub account settings, IDE settings, and project artifacts such as instruction and prompt files.
At the account level, users and administrators can enable or disable features (\eg chat, coding agent, code review, web search), select models, and control data usage.
Within the IDE, developers can select a preferred language and adjust how aggressively it edits code, while project-, path-, and agent-specific instruction files configure coding conventions, frameworks, testing practices, and security rules.
Organizations can add higher-level instructions and content-exclusion rules that apply across repositories, and reusable prompt templates and custom instructions shape how Copilot responds in chat.
Overall, Copilot offers many controls over its behavior, but they are spread across multiple configuration interfaces and are mostly presented as simple toggles or textual prompts.


\textbf{\textit{OpenAI ChatGPT~\cite{chatgpt}.}}
ChatGPT offers configuration options across account-level settings, per-chat controls, project workspaces, custom GPT definitions, and workspace administration panels.
Users can define global custom instructions, enable or disable long-term memory and chat history, and set data-usage and training preferences.
Within a chat, they select models and modes and provide chat-specific instructions, while projects group related chats and reference files under project-specific instructions.
Custom GPTs combine instructions, knowledge files, tools, and sharing settings into reusable assistant profiles, and organizational plans add controls for feature availability, connectors, and data retention.
Overall, ChatGPT’s configurations are relatively focused and easier to locate, since global options are centralized and scope-specific options are attached to the corresponding level, but it offers a more limited set of configurations and does not provide code-specific settings.


\textbf{\textit{Google Gemini~\cite{gemini}.}}
Gemini Code Assist offers configuration options across IDE extension settings, repository-level configuration files, and Google Cloud project settings.
In the IDE, developers can enable or disable inline suggestions, choose supported languages, adjust how and when suggestions appear, and configure local codebase awareness and agent behavior, while repository-level artifacts control which files and patterns should be ignored.
In the Enterprise edition, administrators can connect customization indexes to private repositories so that suggestions draw on internal libraries and standards.
Within the chat interface developers can define custom commands and rules for common workflows.
Overall, Gemini’s coding assistant distributes its configurations across several interfaces and, although it provides a range of code- and system-level configurations, offers little control over end-user personalization and interaction style.


\textbf{\textit{JetBrains AI Assistant~\cite{jbai}.}}
JetBrains AI Assistant offers configuration options across per-project and per-chat IDE settings, repository-level files, and organization-level AI Enterprise controls.
In the IDE, developers can specify where Assistant may modify files without confirmation, choose languages and technologies, adjust disruptive behavior, and connect MCP servers, while quick settings in the chat window expose Ask/Code/Auto modes, a Brave Mode checkbox (autonomy in command execution and file modification), and a model picker.
At the project level, guidelines files provide reusable context about coding conventions, frameworks, testing practices, and examples, while ignore files restrict Assistant’s freedom of action.
Organizational deployments configure which providers and models are available, whether detailed interaction data is collected, and which users can access AI features.
Overall, Assistant’s configurations are concentrated in IDE and project configuration and, while it offers rich options for project scope and code-style guidance, it provides limited support for user-level personalization or cross-project personas.


\textbf{\textit{Cursor~\cite{cursor}.}}
Cursor’s AI-enabled editor offers configuration options across IDE settings, project-level rule and ignore files, and an online team dashboard.
In the IDE, developers can define global user rules for preferred languages, frameworks, code structure, and documentation style, which are applied to chat and inline edits across projects.
Project-level configuration files define repository-specific standards and workflows, while ignore files control which files are indexed and used as context.
System-level controls handle model selection, context size, privacy modes, and across-session memorization, and team dashboards manage enforced privacy settings.
Overall, Cursor has notable code- and system-level control settings but more limited user-personal and task-context controls that are mostly configured indirectly through prompts.


\textbf{\textit{In conclusion.}} Despite growing adoption, the configuration options these tools offer remain fairly limited.
Some allow developers to specify preferred languages or frameworks, while others provide basic system settings, \eg chat memory depth and output size.
However, few offer fine-grained control over the whole range of personalization, including aspects such as behavior, interaction style, company policies, or even color schemes.

Another challenge in today’s assistants’ configuration approach is that the settings that do exist are scattered across multiple poorly connected interfaces.
Developers might find some settings at the account level (\eg data retention, memory preferences), others buried in IDE-specific menus (\eg toggles for inline completions and language activation), and still others in project files (\eg ignore lists that define what the assistant can access) or ad hoc textual prompts and instruction templates.

The result is a fragmented and often confusing configuration experience.
In our survey, a number of developers report being unaware that certain configurations exist (\Cref{surv:results}).
This reflects a wider trend in modern AI systems, where essential functionality becomes distributed across heterogeneous interfaces, making it difficult to discover or manage in a coherent way~\cite{agnia01}.
Configuring the tool becomes cognitively demanding and error-prone rather than straightforward, and this scattered setup makes it hard to build a clear mental model of how the assistant works, which in turn makes it harder to trust or predict its behavior~\cite{amershiDesign}.

Given these challenges, it becomes increasingly important to offer developers configuration mechanisms that are clear, unified, and easy to find.
Prior work in human-AI interaction highlights that user control, predictability, and transparency are key for building and maintaining appropriate trust~\cite{amershiDesign, jennySurvey}.
As AI assistants take on a greater share of daily development work~\cite{DevEco2025, StackoverflowDevSurv2025, jennySurvey}, developers need reliable ways to influence how these systems behave, what constraints they follow, and how they integrate into personal and organizational workflows.

\subsection{Research overview}\label{back:research}
Personalized coding assistants hold substantial potential to increase productivity and trust by aligning with developers needs and expectations~\cite{trustBrown, personalizeLLmsKonda}.
However, implementing automatic personalization poses notable challenges.
According to ~\cite{personalizationTechs}, it requires access to sensitive information, such as private repositories, usage history, or organizational resources, raising concerns about privacy, data retention, and security.
Besides that, complex development workflows, distributed teams, and interdependent codebases might complicate context modeling, limiting the assistant’s ability to reliably generate context-aware suggestions. 
Personalization techniques such as content-based modeling or collaborative filtering may fail in the presence of cold-start scenarios or diverse preference profiles.
That is why many current LLMs mostly rely on prompt-controlled configuration or user-provided feedback for adaptation.

As we described in the previous subsection, most existing assistants offer limited and non-obvious mechanisms for adjusting how they operate, what conventions they follow, or how they interact with the user. 
This one-size-fits-all design has well-documented drawbacks.
Prior work shows that misalignment between a developer’s expectations and the assistant’s behavior can reduce trust, slow adoption, and introduce friction into the development workflow~\cite{trustBrown, cctCustomization, personalizeLLmsKonda}. 
Research in human–AI interaction consistently highlights the importance of adaptability, arguing that AI systems must support users’ goals, context, boundaries, and need for control in order to function effectively~\cite{picse, HAXdesignspace}. 

Prompting is currently one of the most common forms of control over assistant behavior.
According to ~\cite{cursor_prompting}, developers are moving from short-lived, task-specific instructions toward persistent, project-level prompts that are expected to be followed across interactions with the same codebase.
They illustrate this evolution by analyzing 401 open-source repositories with Cursor rules and developing a taxonomy of developer-provided context and directives.
Their findings suggest that developers already treat such rules as a new configuration surface, using them to encode project norms and constrain assistant behavior beyond individual queries.

For developers, adaptability includes not only technical considerations, such as preferred libraries, frameworks, and code style, but also workflow habits, communication preferences, privacy expectations, and organizational constraints.
Having the ability to configure stylistic conventions such as naming schemes, formatting rules, testing structures, or documentation style can help ensure that AI-generated code fits seamlessly into existing codebases and speed up the AI-generation process~\cite{cctCustomization, picse, personalizeLLmsKonda}. 
Similarly, studies of goal-setting and intent-specification interfaces show that allowing users to establish constraints or objectives improves predictability and mutual understanding between humans and AI systems~\cite{contextAIPinto, designForTrustWang}. 
Developers also value control over their interaction with AI, such as timing, verbosity, colors and on-screen positioning, ~\cite{HAXdesignspace, cctCustomization}.

Despite these insights, existing research stops short of identifying \textbf{which} specific configurations developers actually want, \textbf{why} they want them, or \textbf{how} they want to set them up.
This gap is especially noticeable when we look at the general picture and not specific modality, e.g. generated code properties.
In this work, we want to address the gap through structured user research with experienced software developers and outline design and research opportunities for Configurable Coding AI.

\subsection{Compiling a list of configurations}\label{back:compile}
To study developers’ configuration needs in a structured way, we first need an initial, reasonably comprehensive list of configuration options for AI coding assistants. 
This list serves as the basis for our survey and co-design activities, which participants are encouraged to extend with their own suggestions.

We create this list in two steps.
First, we systematically review the configuration settings offered by existing coding assistants, including GitHub Copilot, ChatGPT, Google Gemini, JetBrains AI Assistant, and Cursor (\Cref{back:industry}).
Second, we extend these practical configurations with ideas inspired by prior research on configurable and personalized AI, and trust in AI-powered developer tools (\Cref{back:research}).
This allows us to include not only currently implemented options, but also plausible forward-looking configuration mechanisms that have been proposed in the literature.

We organize our list of configurations into four higher-level groups that reflect what each setting primarily characterizes.
These groups are built in line with prior work on trust in AI-assisted development.
Brown et al.~\cite{trustBrown} identify three types of characteristics that influence trust in AI code completion tools: properties of the suggestion itself, characteristics of the developer, and aspects of the development context. 
The PICSE framework for trust in software tools by Johnson et al.~\cite{picse} groups relevant factors into five categories: Personal, Interaction, Control, System, and Expectations. 
Inspired by these perspectives, we organize our list of configurations into the following groups:
\blue{Characteristics of Code suggestions}, 
\orange{Characteristics of System \& policies}, 
\pink{Characteristics of Human--Assistant interaction}, 
and \purple{Characteristics of Users \& their personal context}.

Our proposed grouping serves two purposes.
Analytically, it provides a manageable structure for comparing how developers prioritize different kinds of configurations, \eg whether they care more about shaping code suggestions or aligning the assistant with their personal context.
Methodologically, it helps us design clear survey blocks, so that participants would think about related configuration options together rather than confronting an undifferentiated list.
Below, we briefly define each group and list the \nConfigs configuration options that we consider in our study.

\blue{\textbf{Characteristics of Code suggestions (CS).}}

These settings are about structure, style and technologies used in the code the assistant suggests: from formatting and language choice to the libraries it uses and how much detail it includes. 

This includes Preferred code style and formatting (\eg tabs or spaces, naming conventions); Restricted APIs, libraries, and files; Preferred programming language(s), versions, and environment constraints (\eg use React with TypeScript rather than plain JS, generate code compatible with Python 3.8); Framework or architecture conventions (\eg functional or OOP, preferred design patterns); Preferred testing style (\eg include unit tests by default, use Jest); Preferred documentation style (\eg generate docstrings, inline comments, or no comments unless asked); and Optimization priorities (\eg readability, performance, or memory).


\orange{\textbf{Characteristics of System \& policies (SP).}}

These settings let the user define preferences related to memory, data handling, execution boundaries, and compliance with policies.

This includes Model selection, Chat memory depth; Memory management and cleanup (\eg delete memorized chats); On-premise vs. off-premise execution; Data retention and privacy preferences (\eg opt-out of logging conversations for training); Suggestion validation (\eg legal license, static analysis, organization style guideline); Organizational governance integration (\eg connect to internal policy engines or shared style guides and linters); User role-based access or permissions (\eg interns' assistants are not allowed to change codebase and access specific files); and Network or data access controls (\eg disable web search, access to internal databases, or external APIs).


\pink{\textbf{Characteristics of Human--assistant interaction (HAI).}}

These settings focus on how the assistant communicates with the user: how it responds, when it responds, and how it looks.

Namely, Preferred response length; Custom shortcuts; Response timing and triggering behavior (\eg only when prompted, frequent auto-completions, snoozed autocompletions); Appearance of code completions (\eg position in editor, opacity or background shading); Showing confidence or quality scores for suggestions; Minimum confidence threshold to show completions; Preferred form of address (\eg by first name, nickname, honorifics, or anonymous); Assigned persona (\eg work buddy, senior expert, review bot, code generation tool); and Preferred communication style (\eg concise or detailed, formal or casual, suggestive or directive).


\purple{\textbf{Characteristics of Users \& their personal context (UPC).}}

These settings focus on the user, their personality, abilities, and work context, so the assistant can adapt to the user’s individual needs and the professional context.

Options here are Gender identity or pronoun preferences; Custom job title or role specification to adjust guidance and expectations (\eg Senior Frontend Engineer, Tech Lead, QA Intern); Accessibility settings (\eg screen reader compatibility, text size, contrast options); Familiarity with AI assistant (\eg novice, experienced user); Business context integration (\eg connect to project management tools like YouTrack, Jira, or GitHub Issues for task-aware assistance); Personal goals or development focus (\eg learning React, improving test coverage habits); Workstyle profiling (\eg through a lightweight personality test or user-selected archetype such as strategist, hacker, careful editor); and Preferred learning style (\eg hands-on guidance or high-level suggestions, step-by-step explanations or summary results).


\section{Survey}\label{survey}

To answer RQ1 and RQ2, we conducted a survey to understand whether and in what ways experienced developers care about configuring AI coding assistants, and to get an initial sense of which configuration options they consider most important.
In this section we report the design of the survey and our findings.

\subsection{Survey Participants}\label{surv:participants}
We recruited participants through an internal research panel at \company, selecting developers who reported at least three years of professional experience and one year of active use of AI coding assistants. 
We intentionally focused on experienced users rather than novices, as they are more likely to have developed stable preferences around assistant behavior and to have encountered concrete configuration needs in real-world practice.
We sent 3,000 invitation emails, receiving 165 link clicks and \nSurvey completed responses. Participants could enter a draw for USD 50 Amazon eGift Cards or an equivalent-value \company product pack.

The participants reported residing in 28 countries, mainly based in Europe ($66\%$). 
$79\%$ are full-time software engineers with substantial professional experience: 15 with 3--5 years, 18 with 6--10 years, 9 with 11--15 years, and 14 with 16+ years. Most had 1--2 years of AI tool experience (about $70\%$), while others had used AI tools for over 2 years.
AI tool use is frequent ($79\%$ daily) and typically IDE-integrated ($89\%$).
The most commonly used tools are ChatGPT ($77\%$), Claude ($48\%$), GitHub Copilot ($46\%$), and JetBrains AI assistant ($38\%$).
Among employed respondents, their company's stance toward AI tools is mixed: 30\% reported that their company encourages use, 27\% allow it with minimal restrictions, 25\% allow it with significant restrictions, and 7\% are uncertain.


\subsection{Survey Outline}\label{surv:outline}

The survey is structured around the list of \nConfigs configurations introduced in ~\Cref{back:compile} and split into four configuration groups: 
\blue{CS}, 
\orange{SP}, 
\pink{HAI}, 
and \purple{UPC}.
The full survey text is available at the Appendix A.1\footnote{https://github.com/KatyaKos/configurableAI}.

The survey starts with background questions that are meant to collect necessary information about the participants and filter out unsuitable candidates.
The filtering criteria are: currently working, at least 3 years of professional experience, a role that involves coding (see question 4 in the survey), and at least 1 year of AI experience.
In the rest of the survey, the suitable candidates are shown one group at a time and asked to evaluate the configurations within that group.
For each configuration option, respondents indicate (1) whether the setting \emph{exists} in the AI tools they currently use (binary: It exists or It doesn’t exist) and (2) how \emph{useful} the setting is or would be for their workflow (3-point scale: Not useful, Moderately useful, Very useful). 
They can optionally add comments and suggest their own configuration ideas that they feel are missing.

\subsection{Survey Findings}\label{surv:results}

In total, \nSurvey participants evaluated \nConfigs configuration options, resulting in 1848 individual ratings.
In general, developers expressed a strong interest in configurability.
$72.6\%$ of all ratings evaluate offered configurations as "Moderately" or "Very useful", which further we call \textbf{positive usefulness}.
When we map these responses onto an ordinal scale ("Not useful" as 0, "Moderately" as 1, "Very" as 2), the overall mean usefulness score is $1.1$ out of 2.
We will further call mean score calculated over such mapping an "ordinal mean".

By contrast, perceived availability lags behind:
only $35.6\%$ of ratings indicate that a given control exists in participants’ current tools.
Together, these results point to a clear configuration gap: developers frequently value controls that they do not believe are currently offered in mainstream AI coding assistants.

Most respondents also reported using configuration to some extent.
$67.9\%$ said that they customize AI tool controls at least occasionally: $17.9\%$ actively customize settings, while another $50\%$ tweak a few options.
At the same time, $8.9\%$ of respondents said that they were not aware that AI assistants can be configured at all, reinforcing concerns about discoverability.

\subsubsection{\textbf{Demand \& Supply}}
We calculated per-configuration summaries and ranked settings to identify the most useful settings, the least useful settings, and settings with the highest unmet demand.
Here, we only provide the top results, the table with all statistics for each configuration can be found in the Appendix A.2.

\paragraph{Highest unmet demand}
We define unmet demand as the difference between positive usefulness rate and existence rate, where larger gaps indicate controls participants value, but do not have or are not aware of their existence.

\circled{1} \pink{\textit{Showing confidence or quality scores for suggestions}} and \pink{\textit{Minimum confidence threshold to show completions}}: $67.9\%$ demand gap.
In comments, some participants explained that, depending on a task, they might want to see more suggestions, even those with lower quality scores.
They mentioned brainstorming as an example of such a task.

\circled{2} \pink{\textit{Preferred response length}}: $62.5\%$ demand gap.
Similarly to the confidence threshold, participants explained that they want to change the verbosity depending on the task context (\eg quick edits vs brainstorming).
  
\circled{3} \orange{\textit{Suggestion validation}}: $55.4\%$ demand gap. 
In the comments, participants mentioned that they want AI tools to "self-check" their output, based on the predefined rules and context.

Overall, unmet demand is highest for the \pink{HAI} group, especially for task-specific controls that influence user trust.

\paragraph{Configurations most frequently perceived useful} \hfill\\
\indent \circled{1} \orange{\textit{Model selection}}: $94.6\%$ positively useful with $82.1\%$ as "Very useful". 
It was also one of the few settings that the participants reported as already available ($91.1\%$)

\circled{2} Both \blue{\textit{Preferred programming language(s), versions, and environment constraints}} and \pink{\textit{Preferred response length}} were rated with $91.1\%$ of positive usefulness. 
However, less than half of the participants identified these controls as existing in their tools.

\circled{3} \orange{\textit{Data retention and privacy preferences}}, \blue{\textit{Preferred testing style}}, \pink{\textit{Showing confidence or quality scores for suggestions}}, and \pink{\textit{Minimum confidence threshold to show completions}} all received $87.5\%$ of positive usefulness score. 
Interestingly, participants identified only \textit{Data retention and privacy preferences} as known to them in their IDEs ($55.4\%$).

\paragraph{Configurations least frequently perceived useful} \hfill\\
\indent \circled{1} \purple{\textit{Gender identity or pronoun preferences}}: $21.4\%$ positively useful and $12.5\%$ knew about this control in their tools.

\circled{2} \purple{\textit{Custom job title or role specification}}: $35.7\%$ positively useful and known to $16.1\%$.

\circled{3} \purple{\textit{Familiarity with AI assistant}}: $46.4\%$ positively useful and known for $17.9\%$.

These results suggest that respondents prioritize configurations that affect correctness, constraints, and trust over purely social or identity-style controls.
Moreover, in the comments one of the participants expressed their discomfort with the \textit{Gender identity} configuration, saying that it is too controversial and unnecessary for work.

\subsubsection{\textbf{Contextual patterns}}
Finally, we analyzed whether participants’ backgrounds have any correlation with how they evaluate configurations.
This analysis was intended to surface trends rather than make causal claims.

\paragraph{AI experience}
We examined whether the perceived usefulness of configurations differs by respondents’ experience using AI tools for coding.
Participants with 1--2 years ($n=39$) of AI use reported a positive usefulness rate of $69.2\%$ and an ordinal mean of $1.02/2$.
Those with more than 2 years ($n=17$) reported $80.6\%$ positive usefulness and a higher ordinal mean of $1.28/2$.
It might suggest that configurability tends to be valued more as AI tools become more embedded in day-to-day work, or that early (2023) adopters are generally more positive towards AI.

\Cref{tab:surv:aiexp-usefulness} breaks down the positive usefulness and ordinal means within each of the four configuration groups.
Interestingly, the rates in the \blue{CS} are very similar between two experience brackets.
Positive usefulness in \orange{SP} is also similar, however, the ordinal means are noticeably higher for more experienced participants, since they voted more configuration options in this group to be "Very useful".
Interestingly, significantly more experienced AI users voted positively in the \pink{HAI} and \purple{UPC} groups, both in terms of positive usefulness and ordinal means.

\paragraph{Company policy}
Respondents in organizations where AI use is actively encouraged ($n=17$) reported a higher overall usefulness (positive $79.9\%$, ordinal mean $1.24$) than those in organizations with significant restrictions (positive $68.4\%$, ordinal mean $0.99$) and minimal restrictions (positive $70.7\%$, ordinal mean $1.11$). 
The per-group statistics are shown in ~\Cref{tab:surv:restrict-usefulness}.

\begin{table}[t]
\centering
\begin{tabular}{lcc}
\toprule
& AI use 1--2 years
& AI use $>$2 years \\
& ($n=39$) & ($n=17$) \\
\midrule
\blue{CS} & 83.9\% | 1.29 & 88.2\% | 1.39 \\
\orange{SP} & 74.9\% | 1.15 & 81.0\% | 1.40 \\
\pink{HAI} & 71.5\% | 1.01 & 85.0\% | 1.27 \\
\purple{UPC} & 47.1\% | 0.67 & 68.4\% | 1.06 \\
\bottomrule
\end{tabular}
\caption{Positive and ordinal mean usefulness rates by AI use experience within configuration groups.}
\label{tab:surv:aiexp-usefulness}
\end{table}

\begin{table}[t]
\centering
\begin{tabular}{lccc}
\toprule
& Encouraged
& Restricts minimal
& Restricts significant \\
& ($n=17$) & ($n=15$) & ($n=14$) \\
\midrule
\blue{CS} & 89.9\% | 1.44 & 76.2\% | 1.19 & 82.7\% | 1.21 \\
\orange{SP} & 84.3\% | 1.31 & 77.0\% | 1.24 & 74.6\% | 1.22 \\
\pink{HAI} & 85.6\% | 1.29 & 71.9\% | 1.13 & 71.4\% | 0.91 \\
\purple{UPC} & 59.6\% | 0.93 & 57.5\% | 0.87 & 45.5\% | 0.61 \\
\bottomrule
\end{tabular}
\caption{Positive and ordinal mean usefulness ratings by organizational policy.}
\label{tab:surv:restrict-usefulness}
\end{table}

\section{Design Sessions}\label{designsess}

\begin{figure}[tpb]
 \includegraphics[width=\linewidth]{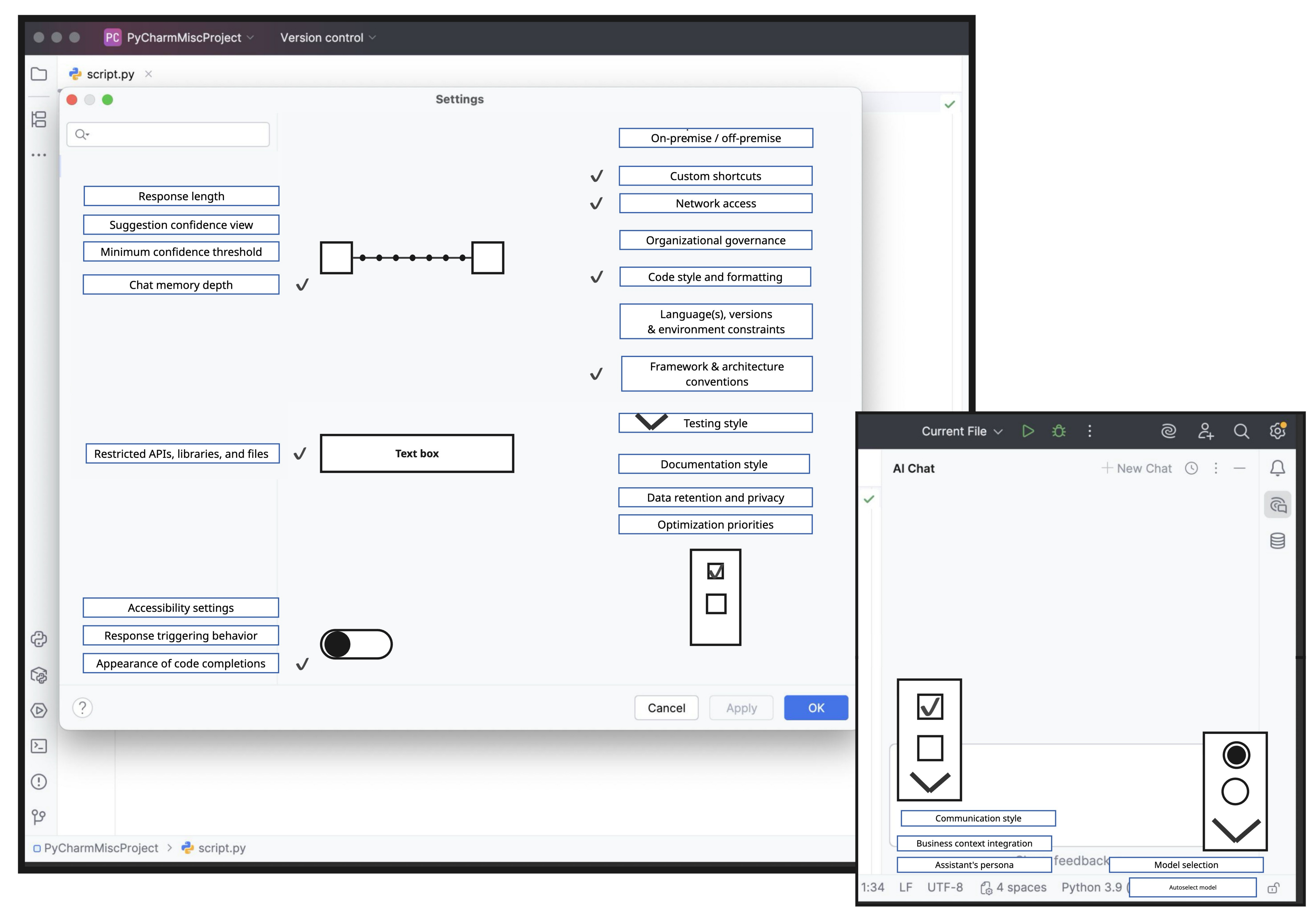}
 \caption{A post-design session Miro board where a participant placed configurations and control widgets (radio button, text box, slider, check box, toggle) either on a main settings window on the left, or a quick-access window on the right. }
 \label{design:main_task}
\end{figure}

To gain more insight into RQ2 and answer RQ3, we conducted design sessions with \nDesign survey respondents who had indicated that they were open to a follow-up interview. 
The goal of these sessions was to move beyond 0-2 rating of individual configuration options and instead explore how experienced developers would actually organize and control settings in an ideal coding assistant: which configurations they would choose to have, where they would place them, which control widgets they would use, and how they reason about priorities and trade-offs.

\subsection{Design Sessions Participants}\label{design:participants}

We conducted \nDesign individual design sessions with professional software developers.
These people were invited as more experienced AI and configuration users out of the survey participants who agreed to be approached further for additional interviews.
All participants reported at least one year of active use of AI coding assistants and five have been using such tools for two years or more.
All participants have substantial professional experience with at least five years of professional software development.
They regularly use tools such as ChatGPT, GitHub Copilot, JetBrains AI Assistant, Gemini, and Claude, typically on a daily basis for tasks like debugging, refactoring, documentation, and learning new technologies.

\subsection{Design Sessions Outline}\label{design:outline}
All \nDesign design sessions were conducted in English, lasted between 60 and 90 minutes, and happened via video call with a shared Miro board. \footnote{https://miro.com/}
The board contained a warm-up task based on familiar IDE settings and the main design task focused on configuring an AI coding assistant.
Participants were asked to think aloud throughout the tasks.

\subsubsection{\textbf{Warm-up}}
We began with a 5-minute task to familiarize participants with the Miro interface and the idea of prioritizing settings.
On the Miro board, participants saw an empty settings screen for PyCharm and a set of cards representing common IDE settings sections: Python, Appearance \& Behavior, Version Control, Languages \& Frameworks, Editor.
Each card had a brief description and examples of the specific controls in that section, extracted from PyCharm.

Participants were asked to drag these cards onto the settings screen and arrange them from most to least important according to any criteria they feel relevant.
We asked them to think aloud about why some settings matter more than others and how they typically approach configuration in their everyday tools.

\subsubsection{\textbf{Main design task}}

After the warm-up, we introduced the main task focused on AI coding assistants (see \Cref{design:main_task} for the finished state of the main task).
The Miro board had two central panels: a larger "settings window" representing a traditional IDE settings dialog, and a narrower "assistant chat window" representing a quick-access area embedded in the assistant’s chat interface.

On the right side, participants saw a list of configuration options, each represented as a draggable card.
These cards corresponded to the configurations from our compiled list (\Cref{back:compile}), presented as a flat list without group labels to avoid biasing participants towards placing configurations from one group close to each other.
The order of items in the list was randomized for each participant.
Participants were instructed to choose which configurations they would want in their ideal coding AI assistant.
We emphasized that they do not have to use each offered option.
If participants felt that an important configuration is missing, they could create a new card using a blank template and label it with their own name and description.
Then they placed each selected configuration in either the main settings window or the quick-access chat panel, depending on where they would prefer to control it in practice.
Participants also attached a control widget to each selected configuration by copying one of the available widgets (toggle, slider, text box, checklist, radio button) and placing it next to the setting card, or by describing another control type they would prefer.

Participants were encouraged to comment on why they place each card in one panel or the other and a certain control widget was chosen, or why they skipped certain configurations and did not place them anywhere.
They were also asked to tell any stories about when they needed such configurations in their work, and voice any ideas on the usage scenarios.

\subsubsection{\textbf{Post-session interviews.}}
After completing the design task, we conducted a short semi-structured interview to find out more about the reasons behind participants’ layouts and control widget choices.
We asked follow-up questions about why certain settings were prioritized or ignored, which controls they expected to adjust most often, and what additional configuration capabilities they wished existed in current tools.
You can see the interview script, together with task explanations, in the Appendix B\footnote{http://github.com/KatyaKos/configurableAI}.

\subsection{Design Sessions Findings}\label{design:results}
\subsubsection{\textbf{How much control?}}
In total, participants made 231 choices across \nDesign sessions and \nConfigs configurations to choose from.
Most of the chosen configurations were placed in the main settings window rather than quick access.
Overall, about $66\%$ of placements assigned configurations to general settings, $14\%$ to quick access, and $20\%$ were left out. 
When we break this down by configuration group, clear patterns emerge.

\blue{\textbf{Code Suggestions.}} 
Configurations that directly affect output were treated as baseline preferences.
Participants almost always placed these in the main settings window ($76\%$), with only a few quick-access placements or leaving them out ($12\%$ each).
All the participants explicitly said that they want to set these up once per project and not change them afterwards.
They also often mentioned configuration files, such as \textit{.yaml, .ignore}, to be a good alternative for the main settings window, 

\orange{\textbf{System \& Policies.}}
System controls turned out to be more debatable in terms of placement (only $62\%$ of placements in main settings).
\textit{On-premise vs. off-premise execution, Data retention and privacy preferences, Suggestion validation, Organizational governance integration, Network or data access} were generally placed in the main settings window, and named as helpful and mostly stable controls that rarely change.
Only two people out of \nDesign said that \textit{User role-based access} could be potentially useful to them, others chose to leave this configuration out.
Six participants placed \textit{Model selection} into a quick-access window, explaining that they all have this configuration in their tools and change it depending on a task they are working on. 
The other one participant mentioned always having the cheapest model on, but expressed interest in a feature that would automatically pick a model based on their task and budget, and three other participants mentioned a similar idea.
Both \textit{Chat memory depth} and \textit{Memory management and cleanup} were split equally between main settings and quick-access windows, with some participants explaining that they would like to control chat message visibility to avoid biasing models with previous dead-end or incorrect reasoning.

\pink{\textbf{Human–Assistant Interaction.}}
While most interaction configuration placements ended up in the main settings window ($72\%$), there were several opinion clashes.
Most notably, half of the participants wanted \textit{Response length} control in main settings, while others requested quick access because, according to one person, AI can be very "bossy" and repetitive, and sometimes needs to be forcefully refocused.
Also, \textit{Preferred form of address} was mostly described by participants as unnecessary.

\purple{\textbf{Users \& Personal Context.}}
User- and personality-centered configurations were rejected the most ($41\%$).
\textit{Business context integration} and \textit{Accessibility settings} were the only controls that got one rejection, with all the others being rejected by at least half of the participants.
When configurations from this group were used, they were always assigned to the main settings window, with the exception of \textit{Business context integration} which was requested as quick-access by two participants in the context of "turning relevant Jira tickets on and off".

All together, these findings show a clear and recurring distinction.
Developers tend to separate consistent project- or organization-specific configurations from task-specific ones.
The former are generally thought to belong in a dedicated settings interface, while the latter are expected to be adjusted more fluidly within the assistant’s interface during everyday use.
One participant suggested that users should be able to build their "control panels" on their own and adjust them to their current needs and tasks.

\subsubsection{\textbf{How to control?}}
One of the design sessions' goals was to see whether particular control widgets would naturally emerge as better fits for specific types of settings.
However, there was rarely a single universally agreed-upon choice.
For most configurations, participants suggested a mix of controls rather than settling on one "correct" widget.
Checkboxes and free-text boxes were the most common choices overall and appeared at least once for nearly every setting.
In many cases, the same configuration was represented with either checkboxes or text boxes by different participants, reflecting a broader split between those who preferred more structured controls and those who favored open-ended input.
We also noticed a strong tendency: most developers implicitly picked a side by leaning either toward check boxes or text box, and then stayed with that choice across most of their configuration decisions.

Still, a certain number of settings attracted near-consensus around specific widget types.
Numeric or continuous parameters, such as \pink{\textit{Minimum confidence threshold to show completions}} and \pink{\textit{Preferred response length}}, were uniformly mapped to sliders.
\pink{\textit{Showing confidence or quality scores for suggestions}} and \orange{\textit{On-premise vs. off-premise execution}} were almost always represented as toggles.
\orange{\textit{Model selection}} was unanimously chosen as a radio button.
Other than this configuration, radio buttons were not a popular choice and were mostly mixed with check boxes and toggles.
The most prominent pattern that we observed was the fact that most \purple{UPC} settings, such as \purple{\textit{Personal goals or development focus}}, were represented with text boxes (when picked).

\subsubsection{\textbf{Additional configuration ideas.}}
Finally, several participants took advantage of the "add your own" option to suggest configurations that were not in our initial list:
\begin{itemize}
  \item Create task-specific profiles (\eg reviewer, explainer, or quick-search mode) for the assistant that could be switched on demand and would have predefined configurations.
  \item Switch for agentic mode
  \item Switch for vibe coding mode
  \item Switch for incognito mode that will not save chats into history or use them for training.
  \item Automatic model selection depending on prompt, resources and overall task
  \item Resource precalculation of the prompt, such as price or CO2 production
\end{itemize}
\section{Discussion}\label{disc}

\textit{\textbf{RQ1: What configurations do experienced developers want?}}

Taken together, the survey and design sessions show a clear interest for configurability in AI coding tools. 
Across all \nConfigs configurations, almost $75\%$ of survey ratings mark options as at least moderately useful, yet just over a third are believed to exist, revealing a clear configuration gap. 
The demand is highest for controls that directly affect trust and task fit.
\textit{\pink{Minimum confidence thresholds}, \pink{Visibility of confidence scores}, \pink{Response length}, \orange{Suggestion validation}} all combine high usefulness with low perceived availability. 
In contrast, social or identity-oriented options, such as \textit{\purple{Gender identity}, \purple{Job titles}, \purple{Workstyle profiling}}, are consistently rated as less important.
Some of the most valued settings, such as \textit{\orange{Model selection}} and \textit{\orange{Data retention preferences}}, are already implemented in mainstream tools, but some participants still described them as hard to find.
Our findings highlight that not only are valued configurations missing, but even existing ones are often fragmented across account pages, IDE panels, project files, and implicit prompting, to the point where developers are unaware they can configure their assistants at all. 

\textit{\textbf{RQ2: How do developers prioritize different types of configuration needs?}}
In the survey, \blue{CS} and \orange{SP} configuration groups stand out as the most important.
We also see from design sessions that participants placed output- and system-related configurations into the main settings window or imagined them living in configuration files, describing them as project or workspace defaults that should be set once and then rarely changed. 
The \pink{HAI} group is close behind and, more interestingly, it contains configurations with the highest unmet demand.
Interaction-level controls were also much more likely to be placed in the assistant’s quick-access area, reflecting their task-specific and situational nature. 
\purple{UPC} options were often rejected entirely, with multiple participants explicitly questioning their relevance to work.
In design sessions, when used at all, they were treated as infrequently touched background settings. 

\textit{\textbf{RQ3: What types of configuration interfaces do developers prefer?}}
The design sessions did not reveal a single dominant control widget, but rather a set of conventions. 
Across configurations, checkboxes and text boxes were the most frequently chosen widgets, and many configurations attracted both, reflecting different preferences for structured versus open-ended control. 
Individual participants tended to stick with one of them, indicating that widget preferences may be as much about personal interaction style as about the configuration itself.

At the same time, some configurations showed near-unanimous choices. 
For example, \textit{\pink{Minimum confidence threshold to show completions}} or \textit{\pink{Response length}} were consistently represented as sliders.
These choices align closely with how current tools already implement similar controls, suggesting that existing interfaces strongly shape developers’ expectations.
In the future, it might prove difficult to move developers from one configuration setup to another, even if the new version proves to be "superior".

\textit{\textbf{Future directions}}
Our participants described some configuration ideas that go beyond the initially proposed list. 
We see them as promising directions for future work, particularly ideas about (1) task-specific assistant profiles, (2) agentic, vibe and incognito mode switches, (3) controls for estimating and constraining per-prompt cost or CO2 footprint.

Finally, our observations hint that configuration needs may interact with user personality and working style. 
Some developers wanted access to as many controls as possible but were happy to bury them deep inside general settings.
Other participants, who also seemed to be more eloquent than others, preferred to set up their preferences through text boxes and configuration files. 
Future research could explicitly investigate how traits such as desired autonomy, risk tolerance, or introversion shape configuration preferences, and how assistants might adapt not only to project context but also to different "configuration personas" among developers.
\subsection{Threats to validity}\label{threats}

\textit{Internal validity.} Our survey relies on self-reported experience, habits, and perceptions of what configurations exist in current tools. 
Participants may misremember which configurations they have seen or used, or respond in ways that are influenced by social desirability.
We partly mitigate this by focusing on experienced assistant users, but we cannot rule out recall or response biases.
In the design sessions, facilitator explanations, example demonstrations, and the structure of the Miro board may also have nudged participants toward particular placements or widget types.
To reduce this risk, we randomized the order of configuration items, avoided showing our four-category grouping, and encouraged participants to add their own settings and controls.

\textit{External validity.}
Our participants were experienced professional developers who already use AI coding assistants regularly.
As a result, the findings may not generalize to novice developers, occasional AI users, or teams working under very different regulatory or organizational constraints.
We see our results as most directly applicable to teams and organizations that are actively (with possible restrictions) integrating AI assistants into development workflows.

\textit{Construct validity.}
We evaluate usefulness via Likert-style ratings but some options may be rated as useful in theory but rarely used in practice.
Our configuration list and four-group structure are also grounded in current tools and prior literature, and therefore cannot cover the full design space.
Finally, the design tasks used a simplified two-panel layout (main settings window and quick-access) and a fixed set of control widgets, which may have constrained how participants expressed their preferences.
We attempted to mitigate this by allowing participants to add free-form settings and to propose alternative controls.

\section{Conclusion}
AI coding assistants are rapidly becoming part of everyday software development, yet their configurability remains limited, fragmented, and often invisible to users.
In this paper, we compiled a list of \nConfigs configuration options for coding assistants, grounded in existing tools and prior work, and organized them into four groups: \blue{Code suggestions}, \orange{System \& policies}, \pink{Human–assistant interaction}, and \purple{Users \& their personal context}.
We then studied how \nSurvey experienced developers and \nDesign design session participants evaluate and structure these options.

Our results show a clear demand gap: most configurations are considered useful in principle, but only a minority are believed to exist in current tools.
Developers particularly want more control over interaction-related aspects such as confidence thresholds, showing confidence scores, and response length, while being skeptical of many persona-related configurations. 
They also implicitly distinguish between stable, project-level controls and more fluid task-specific controls that are expected in quick-access areas.
Together, these findings outline design opportunities for configurable coding assistants and point to the scope for more unified, discoverable, and trustworthy configuration mechanisms.

\section*{Acknowledgements}
We thank all study participants for generously sharing their time, experiences, and insights into their daily development work.
Portions of this manuscript were revised using OpenAI’s ChatGPT for grammar and clarity improvements.
The authors reviewed and edited all suggestions and take full responsibility for the content.


\balance

\bibliographystyle{IEEEtran}
\bibliography{references}

\end{document}